\documentclass{article}
\usepackage{amsmath, amssymb, graphics, setspace}

\newcommand{\mathsym}[1]{{}}
\newcommand{\unicode}[1]{{}}

\begin{document}

Cinem{\' a}tica Topol{\' o}gica

Cristian Valdez

Departamento de F{\' \i}sica de la Universidad del Valle de Guatemala, 18 avenida 11-95 zona 15, 

Vista Hermosa III, Guatemala, Guatemala. 01015.

val08367@gmail.com

\section*{1. Introducci{\' o}n y Motivaci{\' o}n}

En mec{\' a}nica cl{\' a}sica de pregrado nunca hacemos referencia a la curvatura del mundo o del sistema donde nuestra part{\' \i}cula est{\' a}
situada. Por ejemplo, si se imagina una hormiga movi{\' e}ndose sobre una canica, para la hormiga el espacio es muy curvo y no es plano como lo experimenta
la mayor{\' \i}a de nosotros en nuestras actividades cotidianas. Existe un aparato te{\' o}rico para trabajar sobre cualquier \textit{ espacio} de
esta manera, el muy conocido c{\' a}lculo de variedades; sin embargo, un estudiante de pre grado en la Universidad del Valle de Guatemala pocas veces
tiene la oportunidad de cruzarse con las ideas de geometr{\' \i}a diferencial. Entonces, debido a esta situaci{\' o}n resulta importante para mi
intentar reconocer otros posibles m{\' e}todos para comprender la cinem{\' a}tica. Se me ha ocurrido una idea bastante simple para realizar c{\'
a}lculos, quiz{\' a}s nos dejan de ser c{\' a}lculos abstractos, sin embargo deja abierta la posibilidad de estudiar la geometr{\' \i}a de la trayector{\'
\i}a para determinar ciertas propiedades del movimiento. El inicio de esta teor{\' \i}a intenta librarse del esquema direccional de un sistema coordenado.

\section*{2. Descripci{\' o}n Geom{\' e}trica de un Sistema y Movimiento}

Un espacio \(M\) cumple las siguientes condiciones:

$\quad $(S1) \(M\subset \mathbb{R}^3\).

$\quad $(S2) \(M\) es un espacio topol{\' o}gico formado por la topolog{\' \i}a relativa de la topolog{\' \i}a usual de \(\mathbb{R}^3\).

$\quad $(S3) \(M\) es un espacio localmente homeomorfo a \(\mathbb{R}^2\).

$\quad $(S4) \(M\) es un conjunto Lebesgue medible.

Por ejemplo un espacio v{\' a}lido es el siguiente conjunto \(S_2=\left\{(x,y,z)\in \mathbb{R}^3:x^2+y^2+z^2=1\right\}\), la esfera unitaria, en
efecto cuando pensamos en la base abierta de \(S_2\) realmente nos damos cuenta que son los discos abiertos de \(\mathbb{R}^2\) deformados, este
fen{\' o}meno en geometr{\' \i}a diferencial se le conoce como una variedad de dimensi{\' o}n 2, en este sentido la condici{\' o}n 3, significa que
para cada punto en el espacio existe una vecindad del punto que es homeomorfa a \(\mathbb{R}^2\). { }Otro punto importante, la raz{\' o}n por la
cual \(M\) es Lebesgue medible es para que los abiertos de la topolog{\' \i}a relativa sean medibles, i.e., tenemos un sistema de conjuntos medibles
en el espacio \(M\), esto nos abre un mundo infinito de posibilidades. Otros ejemplos de espacios son: cilindros finitos e infinitos, paraboloides,
hiperboloides finitos, cinta de M{\" o}bius, botella de Klein, toroide, entre otras.

Una trayector{\' \i}a \(\zeta :[a,b]\subset \mathbb{R}\to M\subset \mathbb{R}^3\) cumple las siguientes condiciones:

$\quad $(T1) \(\zeta :[a,b]\to M\) es continua

$\quad $(T3) \(\zeta \in C^2\left([a,b],\mathbb{R}^3\right)\)

El intervalo cerrado \([a,b]\) es lo que denominamos como tiempo, adem{\' a}s se supone que \(a<b\). Son dos condiciones simples que tienen un gran
impacto, en primer lugar hay que resaltar que \(\zeta ([a,b])\subset M\), adem{\' a}s \(\zeta ([a,b])\) es compacto, conectado y adem{\' a}s es un
conjunto Lebesgue medible, la trayectoria \(\zeta\) es una funci{\' o}n Lebesgue medible por la continuidad. Una propiedad importante que se estudia
en geometr{\' \i}a diferencial que debe ser mencionada es que el espacio topol{\' o}gico de la imagen de la trayectoria es localmente homeomorfa
a \(\mathbb{R}\), esto significa que \(\zeta ([a,b])\) tiene una estructura topol{\' o}gica similar a un intervalo cerrado de \(\mathbb{R}\); es
como que si hubieramos deformado el intervalo sobre la superficie \(M\). La raz{\' o}n por la cual una trayectoria debe cumplir con ser una curva
suave y derivable dos veces es porque debe satisfacer la ecuaci{\' o}n de Newton, en efecto el operador \(F:C^2\left([a,b],\mathbb{R}^3\right)\to
C\left((a,b),\mathbb{R}^3\right)\), definido como el producto de la \textit{ masa} y la segunda derivada temporal de la trayectoria, nos permite
saber que fuerza genero tal trayectoria, esto raz{\' o}n es una consecuencia del mundo mismo tal y como lo expone V.I. Arnold en su libro \textit{
Mathematical Methods of Classical Mechanics}, no existe forma de demostrar que es cierto el enunciado de una premisa anterior, solamente existen
principios equivalentes.

Consider{\' e} a \(M=S_2\), la esfera unitaria, y \(\zeta (t)=(\cos (t),\sin (t),0)\), con \(t\in [0,2\pi ]\). Este ejemplo claramente establece
las relaciones dadas anteriormente, en efecto, \(M\) es una superficie medible, y \(\zeta :[0,2\pi ]\to M\) es una funci{\' o}n continua y adem{\'
a}s es doblemente diferenciable, donde claramente \(\zeta '(t)=(-\sin (t),\cos (t),0)\) y \(\zeta \text{''}(t)=(-\cos (t),-\sin (t),0)\) son dos
funciones continuas hacia \(\mathbb{R}^3\). Tambi{\' e}n es importante mencionar que \(\zeta ([0,2\pi ])\) es una circunferencia cerrada, compacta,
conectada y medible; b{\' a}sicamente es un conjunto ccon propiedades interesantes.

\section*{3. Cinem{\' a}tica Topol{\' o}gica}

La posici{\' o}n de la part{\' \i}cula est{\' a} dada por \(\zeta (t)\) para un tiempo \(t\), sin embargo que podemos decir acerca de su velocidad.
En mec{\' a}nica cl{\' a}sica sabemos que su velocidad se puede describir como primera derivada temporal de la posici{\' o}n, y adem{\' a}s podemos
completar la funci{\' o}n para que \(\zeta ':[a,b]\to \mathbb{R}^3\), esta manera de calcular la velocidad nos da su comportamiento por componentes.
Esta manera es bastante eficiente de calcular la velocidad, y es la que estamos acostumbrados a usar pero quizas no es lo que necesitamos. Sup{\'
o}ngase una part{\' \i}cula en \(S_2\), y la pregunta ser{\' \i}a {?`}qu{\' e} tan {\' u}til le es a un observador sobre \(S_2\) saber la velocidad
de la part{\' \i}cula respecto de un observador en el origen de \(\mathbb{R}^3\)? Quizas est{\' a} pregunta es complicada de responder, pero este
es un caso simple, sup{\' o}ngase al observador sobre una cinta de M{\" o}bius, ya empieza a ser evidente que realmente se complica para el observador.
{?`}C{\' o}mo puede un observador en una superficie calcular la velocidad? Existe todo un aparato te{\' o}rico para realizar estas operaciones, un
estudiante de pre grado le llama cambio de coordenadas para utilizar coordenadas curvil{\' \i}nias, en el caso de \(S_2\) se utilizan coordenadas
esf{\' e}ricas, sin embargo se quiere evitar el uso de geometr{\' \i}a diferencial.

Sea \(\zeta :[a,b]\to M\) una trayector{\' \i}a, entonces la longitud de arco puede expresarse anal{\' \i}ticamente como:

\(L(\zeta )=\underset{\sqcap \in P([a,b])}{\sup }\sum _{i=1}^n\rho \left(\zeta \left(t_{i-1}\right),\zeta \left(t_i\right)\right)\)

Donde sabemos que \(P([a,b])\) es el conjunto de particiones finitas de la forma \(a=t_0<t_1\text{$\cdots $t}_{n-1}<t_n=b\). En an{\' a}lisis real
se demuestra que la longitud de arco es independiente de la parametrizaci{\' o}n. Esta definici{\' o}n de longitud de arco es simplificada por la
medida de Lebesgue, y en efecto

\(L(\zeta )=m(\zeta ([a,b]))\)

Aqu{\' \i} salta la raz{\' o}n por la cual era necesario que \(M\) fuera un conjunto medible. Para este momento ya es posible obtener la velocidad
promedio de la part{\' \i}cula. Antes de hacer esto realizaremos una complicaci{\' o}n t{\' e}cnica m{\' a}s, sabemos que \([a,b]=\zeta ^{-1}(\zeta
([a,b]))\), en part{\' \i}cular por que \(\zeta\) es continua y sobreyectiva. Entonces podemos definir la rapidez promedio como

\(\langle v\rangle =\frac{m(\zeta ([a,b]))}{m\left(\zeta ^{-1}(\zeta ([a,b]))\right)}\)

En este momento se debe de recordar que la medida de Lebesgue utilizada en el numerador es en \(\mathbb{R}^3\) y la medida de Lebesgue se utiliza
en el cociente es en \(\mathbb{R}\).

Ahora, consider{\' e} \(x\in \zeta ([a,b])\), entonces dado que \(\zeta ([a,b])\) es un subespacio topol{\' o}gico de \(M\), este conjunto tambi{\'
e}n posee una base abierta, como consecuencia que es un sub espacio de \(\mathbb{R}^3\), un espacio localmente conectado; entonces lo anterior implica
que existe un conjunto \(\left\{U_{\lambda }^x\right\}{}_{\lambda \in \Lambda }\) de vecindades conectadas de \(x\) que genera cualquier vecindad
\(V\) de \(x\), las vecindades son conjuntos abiertos conectados, estos conjuntos se pueden imaginar como que fueran de una s{\' o}la pieza.

Para \(x\in \zeta ([a,b])\) seleccione una vecindad \(V\) cualquiera, entonces definimos la rapidez promedio en la vecindad \(V\) como

\(\left\langle v_x,V\right\rangle =\frac{m(V)}{m\left(\zeta ^{-1}(V)\right)}\)

Es importante resaltar en este momento que por la continuidad de \(\zeta\), tenemos que \(\zeta ^{-1}(V)\) es un conjunto abierto de \([a,b]\) y
adem{\' a}s medible. Esta rapidez no tiene direcci{\' o}n, solo tiene magnitud sin embargo es un primer acercamiento para intentar describir la cinem{\'
a}tica sobre un espacio arbitrario, donde resulte casi imposible seleccionar un sistema coordenado adecuado. Ahora para un \(t\in [a,b]\) le corresponde
\(\zeta (t)\), para esta existe un sistema de vecindades \(\left\{U_{\lambda }^{\zeta (t)}\right\}{}_{\lambda \in \Lambda }\). N{\' o}tese la siguiente
relaci{\' o}n, si \(U_{\lambda _i}^{\zeta (t)}\subset U_{\lambda _j}^{\zeta (t)}\), entonces \(m\left(U_{\lambda _i}^{\zeta (t)}\right)\leq m\left(U_{\lambda
_j}^{\zeta (t)}\right)\) y \(m\left(\zeta ^{-1}\left(U_{\lambda _i}^{\zeta (t)}\right)\right)\leq m\left(\zeta ^{-1}\left(U_{\lambda _j}^{\zeta (t)}\right)\right)\),
sin embargo, no necesariamente ocurre que \(\frac{m\left(U_{\lambda _i}^{\zeta (t)}\right)}{m\left(\zeta ^{-1}\left(U_{\lambda _i}^{\zeta (t)}\right)\right)}\leq
\frac{m\left(U_{\lambda _j}^{\zeta (t)}\right)}{m\left(\zeta ^{-1}\left(U_{\lambda _j}^{\zeta (t)}\right)\right)}\). Esto anterior motiva la siguiente
relaci{\' o}n de orden parcial \(\prec\) sobre el siguiente conjunto \(\left\{\left\langle v_{\zeta (t)},U_{\lambda }^{\zeta (t)}\right\rangle :\lambda
\in \Lambda \right\}\), decimos entonces que \(U_{\lambda _j}^{\zeta (t)}\prec U_{\lambda _i}^{\zeta (t)}\) si, y s{\' o}lo si, \(U_{\lambda _i}^{\zeta
(t)}\subset U_{\lambda _j}^{\zeta (t)}\) y \(\frac{m\left(U_{\lambda _i}^{\zeta (t)}\right)}{m\left(\zeta ^{-1}\left(U_{\lambda _i}^{\zeta (t)}\right)\right)}\leq
\frac{m\left(U_{\lambda _j}^{\zeta (t)}\right)}{m\left(\zeta ^{-1}\left(U_{\lambda _j}^{\zeta (t)}\right)\right)}\). Entonces el par ordenado \(\left(\left\{\left\{U_{\lambda
}^{\zeta (t)}\right\}:\lambda \in \Lambda \right\},\prec \right)\) define un conjunto dirigido. Entonces mediante la funci{\' o}n descrita en 1.4,
le llamaremos rapidez vecindal, \(\Sigma :\left\{\left\{U_{\lambda }^{\zeta (t)}\right\}:\lambda \in \Lambda \right\}\to \left\{\left\langle v_{\zeta
(t)},U_{\lambda }^{\zeta (t)}\right\rangle :\lambda \in \Lambda \right\}\) construimos una red, en part{\' \i}cular sabemos que la red converge a
\(\inf \left\{\left\langle v_{\zeta (t)},U_{\lambda }^{\zeta (t)}\right\rangle :\lambda \in \Lambda \right\}\), el cual existe dado que \(\left\{\left\langle
v_{\zeta (t)},U_{\lambda }^{\zeta (t)}\right\rangle :\lambda \in \Lambda \right\}\) est{\' a} acotado inferiormente y es no vac{\' \i}o. En base
el tema expuesto anteriormente definimos la rapidez instantan{\' e}a para \(t\) como

\(v(t)=\inf \left\{\left\langle v_{\zeta (t)},U_{\lambda }^{\zeta (t)}\right\rangle :\lambda \in \Lambda \right\}\)

De igual manera se puede trabajar con la aceleraci{\' o}n utilizadon la funci{\' o}n extendida de la velocidad como la primera derivada temporal
de la trayectoria, i.e., { }realizando el proceso descrito anteriormente para \(\zeta '(t):[a,b]\to \mathbb{R}^3\). Entonces utilizando la topolog{\'
\i}a de \(\mathbb{R}^n\) se ha construido una teor{\' \i}a de cinem{\' a}tica diferente a la usual, en part{\' \i}cular porque nos hemos salido del
concepto de considerar las trayectorias como objetos funcionales, sino como objetos geom{\' e}tricos.

\section*{4. Referencias}

T. Terence, \textit{ Analysis II}, \textit{ 1}th ed., India: Hindustan Book Agency, 2006.

A.N. Kolmogorov y S.V. Fomin, \textit{ Introductory Real Analysis}, \textit{ 1}th ed., United States: Dover Publications, 1975.

M. Gemignani, \textit{ Elementary Topology}, 2nd ed., United States: Dover Publications, 1992

V.I. Arnol{'}d, \textit{ Mathematical Methods of Classical Mechanics, }2nd ed., United States: Springer-Verlag New York, 1989.

K. J{\" a}nich, \textit{ Topology}, 2nd. ed., United States: Springer-Verlag New York, 1980.

\end{document}